\documentclass[a4paper]{article}

\usepackage[includeheadfoot, left=3cm, right=3cm, top=2cm, bottom = 2cm]{geometry}
\usepackage{fancyhdr}
\usepackage{authblk}

\usepackage[utf8]{inputenc}
\usepackage{amsmath}
\newtheorem{definition}{Definition}

\usepackage{amsfonts}
\usepackage{amssymb}
\usepackage[hidelinks]{hyperref}
\usepackage{orcidlink}
\usepackage{doi}
\usepackage{xcolor}
\usepackage{tabularx}
\usepackage{zed-csp}

\usepackage{adjustbox}

\usepackage{paralist}

\usepackage[xindy]{glossaries}

\usepackage{xspace}

\newcommand{\Varanus}{\textsc{Varanus}\xspace}

\newcommand{\Rvbip}{RV-BIP\xspace}

\newcommand{\assert}{\textit{ASSERT}$^{\text{TM}}$}

\newacronym{ltl}{LTL}{Linear-time Temporal Logic}

\newacronym{foltl}{FOLTL}{First-Order Linear-time Temporal Logic}

\newacronym{ctl}{CTL}{Computation Tree Logic}

\newacronym{ptl}{PTL}{Probabilistic Temporal Logic}

\newacronym{pctl}{PCTL}{Probabilistic Computation Tree Logic}

\newacronym{pctl*}{PCTL*}{Probabilistic Computation Tree Logic*}

\newacronym{tctl}{TCTL}{Timed Computation Tree Logic}

\newacronym{stl}{STL}{Signal Temporal Logic}

\newacronym{ltl-x}{LTL-X}{a version of Linear Time Logic without the `next' operator}

\newacronym{mtl}{MTL}{Metric Temporal Logic}

\newacronym{iactlk-x}{IACTLK-X}{a fragment of the temporal-epistemic logic CTLK, without the `next' operator}

\newacronym{mucalc}{$\mu$-calculus}{a strict superset of other temporal logics}

\newacronym{fotl}{FOTL}{First-Order Temporal Logic}

\newacronym{bdi}{BDI}{Belief-Desire-Intention}

\newacronym{prs}{PRS}{Procedural Reasoning System}

\newacronym{are}{ARE}{Autonomous Reasoning Engine}

\newacronym{dmars}{dMARS}{distributed Multi-Agent Reasoning System}

\newacronym{mas}{MAS}{Multi-Agent System}

\newacronym{asm}{ASM}{Abstract State Machines}

\newacronym{tasm}{TASM}{Timed Abstract State Machines}

\newacronym{fsm}{FSM}{Finite-State Machine}

\newacronym{pfsm}{PFSM}{Probabilistic Finite-State Machine}

\newacronym{apfsm}{APFSM}{\textit{Autonomous} Probabilistic Finite-State Machine}

\newacronym{pha}{PHA}{Probabilistic Hybrid Automata}

\newacronym{fsa}{FSA}{Finite-State Automata}

\newacronym{ta}{TA}{Timed Automata}

\newacronym{pta}{PTA}{Probabilistic Timed Automata}

\newacronym{gspn}{GSPN}{Generalised Stochastic Petri Net}

\newacronym{mopn}{MOPN}{Marked Ordinary Petri Net}

\newacronym{cpn}{CPN}{Coloured Petri Net}

\newacronym{ba}{BA}{B\"{u}chi Automata}

\newacronym{dtmc}{DTMC}{Discrete-Time Markov Chain}

\newacronym{ctmc}{CTMC}{Continuous-Time Markov Chain}

\newacronym{mdp}{MDP}{Markov Decision Process}

\newacronym{ioa}{IOA}{Input/Output Automata}

\newacronym{pars}{PARS}{Process Algebra for Robot Schemas}

\newacronym{fsp}{FSP}{Finite State Processes}

\newacronym{piadl}{$\pi$ADL}{$\pi$ calculus combined with ADL}

\newacronym{csp}{CSP}{Communicating Sequential Processes}

\newacronym{hcsp}{HCSP}{Hybrid Communicating Sequential Processes}

\newacronym{shcsp}{SHCSP}{Stochastic Hybrid Communicating Sequential Processes}

\newacronym{ccs}{CCS}{Calculus of Communicating Systems}

\newacronym{wsccs}{WSCCS}{Weighted Synchronous CCS}

\newacronym{csp|b}{CSP$\|$B}{an integrated formalism combining CSP and B}

\newacronym{vhdl}{VHDL}{VHSIC Hardware Description Language}

\newacronym{adl}{ADL}{Architecture Description Language}

\newacronym{slim}{SLIM}{System Level Integration Modelling}

\newacronym{mal}{MAL}{Model Action Logic Language}

\newacronym{acsl}{ACSL}{the ANSI C Specification Language}

\newacronym{fdr}{FDR}{the Failures-Divergences Refinement checker}

\newacronym{jpf}{JPF}{Java PathFinder}

\newacronym{ajpf}{AJPF}{Agent Java PathFinder}

\newacronym{mcmas}{MCMAS}{Model Checker for Multi-Agent Systems}

\newacronym{ltlmop}{LTLMoP}{Linear Temporal Logic MissiOn Planning}

\newacronym{cbmc}{CBMC}{C Bounded Model Checker}

\newacronym{cadp}{CADP}{Construction and Analysis of Distributed Processes}
\newacronym{spear}{SpeAR}{Specification and Analysis for Requirements Tool}

\newacronym{psl}{PSL}{Property Specification Language}

\newacronym{fpga}{FPGA}{Field-Programmable Gate Array}

\newacronym{aadl}{AADL}{Architecture Analysis and Design Language}

\newacronym{lts}{LTS}{Labelled-Transition System}

\newacronym{uml}{UML}{Unified Modelling Language}

\newacronym{sysml}{SySML}{Systems Modelling Language}

\newacronym{robotml}{RobotML}{Robot Modelling Language}

\newacronym{dsl}{DSL}{Domain Specific Language}

\newacronym{sct}{SCT}{Supervisory Control Theory}

\newacronym{sat}{SAT}{Boolean satisfiability problem}

\newacronym{smt}{SMT}{Satisfiability Modulo Theory}

\newacronym{ide}{IDE}{Integrated Development Environment}

\newacronym{mde}{MDE}{Model-Driven Engineering}

\newacronym{ai}{AI}{Artificial Intelligence}

\newacronym{fdir}{FDIR}{Fault Detection, Isolation and Recovery}

\newacronym{ifm}{iFM}{Integrated Formal Methods}

\newacronym{ros}{ROS}{Robot Operating System}

\newacronym{amcl}{AMCL}{Adaptive Monte Carlo Localisation}

\newacronym{dl}{\text{\upshape\textsf{d{\kern-0.05em}L}}}{differential dynamic logic}

\newacronym{vm}{VM}{Virtual Machine}

\newacronym{bip}{BIP}{Behaviour Interaction Priority}

\newacronym{fol}{FOL}{First-Order Logic}

\newacronym{owl}{OWL}{Web Ontology Language}

\newacronym{ria}{RIA}{Restore Invariant Approach}

\newacronym{ocl}{OCL}{Object Constraint Language}

\newacronym{sil}{SIL}{Safety Integrity Level}

\newacronym{rcl}{RCL}{ROS Contract Language}

\newacronym{rv}{RV}{Runtime Verification}

\newacronym{rml}{RML}{Runtime Monitoring Language}

\newacronym{sats}{SATS}{Small Aircraft Transportation System}

\newacronym{ico}{ICO}{Interactive Cooperative Objects}

\newacronym{gui}{GUI}{Graphical User Interface}

\newacronym{can}{CAN}{Controller Area Network}

\newacronym{pals}{PALS}{Physically Asynchronous, Logically Synchronous}

\newacronym{cps}{CPS}{Cyber-Physical System}

\newacronym{cots}{COTS}{Commercial Off-The-Shelf}

\newacronym{mtbf}{MTBF}{Mean Time Between Failures}

\newacronym{wcet}{WCET}{Worst-Case Execution Time}

\newacronym{sua}{SUA}{System Under Analysis}

\newacronym{caa}{CAA}{Civilian Aviation Authority}

\newacronym{atc}{ATC}{Air Traffic Control}

\newacronym{dlr}{DLR}{German Aerospace Centre}

\newacronym{assert}{\assert}{Analysis of Semantic Specifications and Efficient generation of Requirements-based Tests}

\newacronym{rae}{RAE}{Requirements Analysis Engine}

\newacronym{esa}{ESA}{European Space Agency}

\usepackage[
    type={CC},
    modifier={by-nc-sa},
    version={4.0},
]{doclicense}

\fancypagestyle{firststyle}
{  
   \fancyhf{}
   \fancyfoot[L]{
   \begin{minipage}{\textwidth}
   \centering {\footnotesize \doclicenseThis }
   \thepage 
\end{minipage}  
}
 
}

\begin{document}

\title{Varanus: Runtime Verification for CSP}

\author[1]{Matt Luckcuck \orcidlink{0000-0002-6444-9312}}
\author[2]{Angelo Ferrando \orcidlink{0000-0002-8711-4670}}
\author[3]{Fatma Faruq}
\date{\today}

\affil[1]{School of Computer Science, University of Nottingham, UK}
\affil[2]{University of Modena and Reggio Emilia, Italy}
\affil[3]{Independent Researcher, UK }

\maketitle 
\thispagestyle{firststyle}

\begin{abstract}

Autonomous systems are often used in changeable and unknown environments, where traditional verification may not be suitable. 
\gls{rv} checks events performed by a system against a formal specification of its intended behaviour, making it highly suitable for ensuring that an autonomous system is obeying its specification at runtime. 
\gls{csp} is a process algebra usually used in static verification,  which captures behaviour as a trace of events, making it useful for \gls{rv} as well. Further, \gls{csp} has more recently been used to specify autonomous and robotic systems. 
Though \gls{csp} is supported by two extant model checkers, so far it has no \gls{rv} tool.
This paper presents \Varanus, an \gls{rv} tool that monitors a system against an oracle built from a \gls{csp} specification. This approach enables the reuse  without modifications of a specification that was built, e.g during the system's design. We describe the tool, apply it to a simulated autonomous robotic rover inspecting a nuclear waste store, empirically comparing its performance to two other RV tools using different languages, and demonstrate how it can detect violations of the specification. \Varanus can synthesise a monitor from a \gls{csp} process in roughly linear time, with respect to the number of states and transitions in the model; and checks each event in roughly constant time.

\end{abstract}

\glsresetall

\section{Introduction}
\label{sec:intro}

Robotic and Autonomous Systems (RAS) must operate reliably in unpredictable, dynamic environments, where design-time assumptions may not hold and traditional verification techniques often fall short. \gls{rv} addresses this challenge by comparing events performed by a \gls{sua} to a formal specification of its intended behaviour; this specification, or \textit{oracle}, determines whether the \gls{sua} has complied with it. \gls{rv} can be performed \textit{online}, on a running program, or \textit{offline}, by analysing a logged execution trace. It enables the behaviour of an autonomous \gls{sua} to drive exploration of the specification’s state space, providing ongoing assurance during operation and bridging the \textit{reality gap} between design-time models and real-world execution~\cite{luckcuck_using_2021}.

This paper presents \Varanus, an \gls{rv} tool that meets the challenges of autonomous systems. It uses a \gls{csp} specification as its oracle. Although \gls{csp} is a well-established process algebra with two model checkers -- \gls{fdr}~\cite{fdr} and the Process Analysis Toolkit~\cite{pat} -- it previously lacked dedicated support for \gls{rv}. Though this work focusses on autonomous systems, we note that \Varanus can also be used for the \gls{rv} of non-autonomous systems.

\gls{csp} is a rich, flexible language often used to model requirements or high-level designs, and has more recently been used to specify autonomous and robotic systems~\cite{Miyazawa2019}. Specifications at varying abstraction levels can be linked through refinement, and \gls{csp} enables reasoning about concurrency, synchronisation, and message-passing. \Varanus synthesises its oracle directly from a \gls{csp} process, in contrast to approaches that rely on \gls{csp} implementations or dialects (see \S\ref{sec:relatedWork}).

A key advantage of our approach is that \Varanus verifies behaviour against a detailed process model, not just individual properties. This enables the reuse of formally verified specifications from earlier development stages and supports validating existing \gls{csp} models against implementations, closing the verification loop. 
Reusing a verified model lends confidence that the oracle is correct, and helps amortise the time spent building the model~\cite{rozier2016}.

RAS are typically component-based (e.g., \gls{ros}) and involve complex communication patterns. \gls{csp} is particularly suited to this context, allowing modular specifications and supporting detailed reasoning about message ordering. \Varanus leverages these features to detect violations such as incorrect sequencing or mismatched communication.

We demonstrate \Varanus using a simulated autonomous rover patrolling a nuclear waste store, monitored against a \gls{csp} specification that captures the minimal behaviour required to meet four system requirements. We evaluate \Varanus by checking traces from this simulation against the \gls{csp} model.

We also stress-test \Varanus using randomly generated \gls{csp} models and traces of increasing size to empirically show its linear performance. Additionally, we compare \Varanus’ synthesis times with two state-of-the-art \gls{rv} tools. While our evaluations are performed offline, \Varanus also supports online \gls{rv}; however instrumenting this case study for online use proved difficult, so this is left as future work. Nonetheless, we show that the \gls{csp} oracle can be synthesised in roughly linear time, and each event checked in roughly constant time (see \S\ref{sec:evaluation}).

The rest of the paper is structured as follows: \S\ref{sec:relatedWork} covers related work; \S\ref{sec:varanus} introduces \gls{csp} and \Varanus; \S\ref{sec:caseStudy} presents the rover case study; \S\ref{sec:evaluation} reports the evaluation; and \S\ref{sec:conclusion} concludes with future work.

\section{Related Work}
\label{sec:relatedWork}

Many robotics \glspl{dsl} prioritise code generation or simulation over verification~\cite{Nordmann2016}, highlighting the need for verification in robotic systems. Our survey on Formal Methods for Autonomous Robotic Systems~\cite{luckcuck_formal_2019} identifies a notable exception: the \gls{bip} approach, which supports both specification and verification~\cite{basu2008incremental,basu2011rigorous,bensalem2009designing}, and has been extended with \Rvbip{}~\cite{falcone2011runtime}. More recently, RoboChart~\cite{Miyazawa2019} offers a diagrammatic modelling language with a (timed) tock-CSP semantics; the structure and timing constructs provided by RoboChart models are a future target for \Varanus. 

Our approach uses a \gls{csp} model directly as the \gls{rv} oracle. Cavalcanti et al.~\cite{cspAndKripke} proposed translating divergence-free \gls{csp} processes into Kripke structures for \gls{rv}, but no tool has resulted from this work. Other approaches rely on \gls{csp} dialects or implementations. For example, the \textbf{\textsf{Jass}} system~\cite{bartetzko_jass_2001} includes an assertion for permissible method traces using a \gls{csp} dialect tailored to Java~\cite{moller_specifying_2002}, limiting its general reuse. Our work is agnostic of the \gls{sua}'s implementation. Similarly, $CSP_E$~\cite{yamagata_runtime_2016}, a Scala-based shallow-embedded \gls{dsl}, requires converting models and lacks full \gls{csp} support, limiting its reusability.

Dynamic verification has also been applied to mobile robotics, such as predicting collisions with humans~\cite{Liu2017} or safe robot arm operation~\cite{Pereira2015}. These are online methods, but do not provide static validation.

By contrast, ModelPlex~\cite{Mitsch2016} combines offline verification of hybrid models in \gls{dl} with online validation, triggering safe behaviour when deviations occur. Another approach checks runtime behaviour of an autonomous robot against a model of its deployment environment~\cite{DBLP:journals/tosem/FerrandoDCFAM21}, revealing where real-world interactions break design-time assumptions. While similar in goal, our case study is human-controlled and does not model its environment.


Rule-based systems are popular in \gls{rv} for efficient event matching, particularly those using the Rete algorithm~\cite{DBLP:journals/ai/Forgy82}. LogFire~\cite{DBLP:journals/sttt/Havelund15}, a Scala-based tool, and Drools\footnote{\url{https://www.drools.org/}}, a widely-used rule engine, use Rete to minimise re-evaluation by structuring rules as networks. These frameworks are efficient and flexible, by contrast \Varanus avoids rule-centric design by using \gls{csp}.

Temporal logic-based \gls{rv} tools, such as LogScope~\cite{DBLP:journals/jacic/BarringerGHS10} and TraceContract~\cite{DBLP:conf/fm/BarringerH11}, support complex event monitoring via parameterised temporal logic. While expressive, they may struggle with performance on large event sets. \Varanus offers a more scalable alternative by modelling concurrency and dependencies directly in \gls{csp}, enabling efficient handling of complex sequences.

The Monitoring Oriented Programming (MOP) framework~\cite{DBLP:journals/sttt/MeredithJGCR12} improves efficiency through parametric trace slicing, creating independent traces per event parameter. In contrast, \Varanus achieves linear performance without slicing, offering a simpler solution aligned with CSP models.

Complex Event Processing (CEP)~\cite{DBLP:conf/ruleml/Luckham08} targets real-time event pattern recognition in large data streams. Drools integrates CEP-like features for adaptive rule processing. However, \Varanus takes a deterministic approach focused on structured event sequences, making it better suited for ordered process models.

Finally, \gls{rml}~\cite{DBLP:journals/scp/AnconaFFM21} specifies monitors with complex temporal and state-based conditions. Unlike \Varanus, which uses \gls{csp} directly, \gls{rml} emphasises rule-based and state-machine monitoring. \Varanus targets structured verification in concurrent and distributed systems, where strict event ordering and concurrency control are essential. Its model-driven approach complements \gls{rml}'s flexibility, offering robust support for high-assurance environments.


\section{Varanus: CSP Runtime Verification}
\label{sec:varanus}

This section describes \Varanus, our \gls{csp} verification tool\footnote{{\Varanus}, named for the biological genus of Monitor Lizards, is available at \url{https://github.com/autonomy-and-verification/varanus/releases/tag/v0.9.4}.}.
\Varanus is capable of monitoring the \gls{sua} against a \gls{csp} process that specifies its intended behaviour.
While the \gls{csp} model captures only the intended behaviour of the \gls{sua}, not its environment, failures at runtime can expose where the environment does not behave as the \gls{sua} assumes.
 The current version of \Varanus (v0.9.4) uses \gls{fdr}'s API to parse the \gls{csp} specification and to check if the \gls{csp} process is suitable, given the events in the \gls{sua} that are visible, via a determinism check (described in \S\ref{sec:varanusOverview}). We aim to remove the dependency on \gls{fdr} and limitation of the tool to deterministic processes in future work. 

The \gls{sua} can be monitored \textit{online}, where \Varanus listens for events performed by a running \gls{sua}; or \textit{offline}, where \Varanus reads events from a logged trace of the \gls{sua}'s behaviour. 
Our evaluation focusses on offline \gls{rv}. In either case, \Varanus checks if the event is valid, given the current trace of events and returns a verdict (pass or fail); for a failing trace, \Varanus will warn the user, then display the failing trace and acceptable events (a counterexample). 
In the rest of this section, we briefly describe \gls{csp} (\S\ref{sec:cspOverview}) to give the reader an understanding of the language used to specify the system's intended behaviour, and give an overview of how \Varanus works (\S\ref{sec:varanusOverview}).

\subsection{CSP Overview}
\label{sec:cspOverview}

\begin{table*}[t]

\caption{Summary of \gls{csp} operators used in this paper. \label{tab:cspOperators}}

\centering
\begin{tabularx}{\textwidth}{ l | l | X  }
    \hline
\textbf{Action}			& \textbf{Syntax} 				& \textbf{Description} \\
\hline \hline
Skip				& $\Skip$ 				& The terminating process \\
\hline
Simple Prefix	&  $a \then \Skip$		& Simple synchronisation on $a$ with no data, followed by $\Skip$ \\
\hline
Input Event 		& $a?in$					& Synchronisation that binds a the input value to $in$ \\
\hline
Restricted Input Event & $ a?in:(valid)$ & Restricts the possible values of $in$ to members of $valid$ \\
\hline
Parameter Event	& $c.value$				& Synchronisation matching the given $value$ \\
\hline
Guarded Event  &  $(boolean~expression)\& c.var$ & Guards communication on $c$ with a boolean expression \\
\hline
External Choice & $P \extchoice Q$		& Choice between two processes $P$ and $Q$\\
\hline \hline
\end{tabularx}

\end{table*}

This section provides a brief overview of \gls{csp} as a primer for the reader. Table~\ref{tab:cspOperators} shows the operators that we use in this paper.
\gls{csp} specifications are built from processes, which may take parameters. A process describes a sequence of events; for example $a \then b \then \Skip$ is the process where the events $a$ and $b$ happen sequentially, followed by $\Skip$ which is the terminating process. 
An event is an instantaneous communication on a \textit{channel}, and each event marks something that we are interested in during the system's execution.
 Channels enable message-passing between processes, they are synchronous, non-lossy, and may have multiple end-points; but a process may also perform an event (communicate an event on a channel) independently without a cooperating process. 

A channel may declare typed parameters, for example {$channel~c~:~int$} declares a channel $c$ with one integer parameter. Parameters communicated on $c$ may be inputs ($c?in$), outputs ($c!out$), or a given value ($c.value$); here $in$, $out$, and $value$ are all of type $int$ to match the parameter type. Input parameters bind to the parameter name, and output parameters must already exist in the scope of the process. Inputs can be restricted ($c?p:(set)$) to only parameters ($p$) in a given $set$ -- since $c$ takes an $int$ parameter, $set$ is a set of $int$s. 
$P \extchoice Q$ offers the alternative of either $P$ or $Q$, once the first event of either $P$ or $Q$ occurs, the other becomes unavailable. 

\gls{fdr}'s input language, \texttt{$CSP_M$}, is a machine-readable version of \gls{csp} that adds basic set functions like $member(e, S)$, which returns true if $e$ is a member of the set $S$; and $diff(S, T)$, which returns the set $S$ with the elements from set $T$ removed. For verification, \gls{fdr} has in-built assertions and can check \textit{refinement} between processes. The in-built assertions are used to check for \textit{deadlock}, where a process can no longer perform any events; \textit{divergence}, where a process may perform an infinite sequence of internal events; and \textit{non-determinism}, where a process can both accept and refuse to perform an event after some initial trace. A refinement assertion compares the behaviour of the processes. For example, if we have two processes $P$ and $Q$, then $P$ is refined by $Q$ ($ P \sqsubseteq Q$) if every behaviour of $Q$ is also a behaviour of $P$. This can be thought of as $Q$ implementing $P$, like a software component implementing an interface.

\subsection{Varanus Overview}
\label{sec:varanusOverview}

\Varanus provides a terminal interface and most of its options can be set using a YAML configuration file.  
When we are confident that the \gls{csp} specification accurately represents the system's intended behaviour, we can use it for \gls{rv}. 
As previously mentioned, \Varanus uses \gls{fdr} to parse \gls{csp} specification, explore the \gls{lts} produced by \gls{fdr} to synthesise our \gls{rv} oracle (which is also an \gls{lts}). 
 We do not directly use the data structure produced by \gls{fdr} because its own documentation says that ``the methods that are used to visit transitions of this machine are not high-performance''\footnote{\url{https://cocotec.io/fdr/manual/api/c++/class_f_d_r_1_1_l_t_s_1_1_machine.html}}. 
\Varanus assumes that the \gls{csp} specification can be correctly parsed by \gls{fdr} and has been verified as a correct specification of the intended behaviour of the \gls{sua}. 

To monitor the \gls{sua}, \Varanus processes each event performed by the \gls{sua}, either by reading a file of logged events or listening over sockets/WebSockets.
The information/events that \Varanus can read is, clearly, dependent on the \gls{sua} itself. But if the event names do not match the events in the \gls{csp} model, then \Varanus can read a JSON file that contains mappings between the \gls{sua} events and the \gls{csp} events. This must be written manually, because it is highly dependent on the \gls{sua} and specification. 

\Varanus checks the behaviour of the \gls{sua} against the formal model being used as the oracle. In this formal model, $M$, a transition from one state to another is labelled by the event that triggers that transition; the set of all the events that the model can perform is called its alphabet, and we use the function $Alpha(M)$ to get the alphabet of model $M$. The \gls{sua} can be thought of as a `hidden' model, the transitions of which are revealed one event at a time. 

For each new event, $e$, produced by the \gls{sua}, \Varanus checks if $e$ is the label of one of the transitions leaving the current state; we represent this set of events using the function $Events(s)$, which produces the events of the transitions leaving state $s$. If the event is in the model's alphabet ($e \in Alpha(M)$) and in the set of event labels for the current state ($e \in Events(s)$) then the event is valid and the current state in the \gls{lts} moves to the transition's destination state, $s'$.

If the event is in the model's alphabet but is not available in this state ($e \in Alpha(M)$ but $e \notin Events(s)$), then \Varanus will always conclude that this is a failing trace of events. However, if the event is not in the model's alphabet ($e \notin Alpha(M)$ and, presumably, $e \notin Events(s)$ either) then we have a problem, because we don't know why the alphabets of the model and \gls{sua} are different. Maybe:
\begin{compactenum}
\item the model is complete and the \gls{sua} is misbehaving; or, 
\item the model is partial and the correctness of the \gls{sua} does not depend on $e$.
\end{compactenum}
So \Varanus provides two modes, \textit{strict} and \textit{permissive}, to deal with these two possibilities.  Strict mode handles situation 1, above; if \Varanus receives an event that is not in the model's alphabet, then it sees this as an error in the \gls{sua}.  
Permissive mode handles situation 2: if \Varanus receives an event not in the alphabet, then it simply stays in the current state. 

In strict mode, we assume that the \gls{csp} model is complete, it captures exactly what the system should do. In this mode, if an event from the \gls{sua} is not in $Alpha(M)$, then \Varanus reports an error because the model does not have a corresponding transition, therefore the \gls{sua} must be misbehaving. Definition~\ref{def:nextStrict} shows the next-state function for \Varanus's strict mode.

\begin{definition}[\textbf{Next-State in Strict Mode}]    
\label{def:nextStrict}
\[
 next_{strict}(s, e) =
  \begin{cases} 
   Error & \text{if } e \notin Alpha(M) \\
   Error & \text{if } e \notin Events(s) \\
   s' & \text{otherwise } 
  \end{cases}
\]

\end{definition}

In permissive mode we assume that the model is an patrial, or abstract specification of what the system should do. Here, if an event from the \gls{sua} is not in $Alpha(M)$, then \Varanus ignores it and stays in the current state. Definition~\ref{def:nextPermissive} shows the next-state function for permissive mode.

\begin{definition}[\textbf{Next-State in Permissive Mode}]
\label{def:nextPermissive}
\[
 next_{permissive}(s, e) =
  \begin{cases} 
   s & \text{if } e \notin Alpha(M) \\
   Error & \text{if } e \notin Events(s) \\
   s' & \text{otherwise } 
  \end{cases}
\]
\end{definition}

For offline \gls{rv}, \Varanus processes each event from the logged events in the trace file, checking it and transitioning to the next state, until it either reaches the end of the trace file or finds a failing event.  For online \gls{rv}, \Varanus listens for, and checks, events until the \gls{sua} disconnects or it finds a failing event.
In both cases, if a new event produces a trace that fails, \Varanus will report this to the user. In future work, we want to improve the interaction between \Varanus and the \gls{sua} to enable runtime \textit{enforcement}, where behaviour that violates the specification can be corrected. We discuss this more in \S\ref{sec:conclusion}. 

A current limitation of \Varanus (v0.9.4) is that it can only use deterministic \gls{csp} processes, this is to simplify how \Varanus deals with a mismatch between the alphabets of the \gls{csp} model and the \gls{sua}; we aim to remove this limitation in future work. 
This mismatch causes a problem when the alphabet of the \gls{csp} model is a larger than the \gls{sua}; \gls{csp} lets us hide the events that the model can perform but the \gls{sua} cannot, but these hidden events can cause non-determinism which makes it hard to decide if the \gls{sua} has violated the specification or not. \Varanus checks that the \gls{csp} process is deterministic, while hiding the events that the \gls{sua} cannot perform. If the process is non-deterministic, \Varanus aborts before even synthesising the oracle.

\section{Case Study}
\label{sec:caseStudy}

Our case study involves an autonomous ground rover patrolling a simulated nuclear waste store~\cite{luckcuck2023compositional}. The rover must inspect five waypoints and take radiation readings -- categorised as \textit{red} (high), \textit{orange} (medium), or \textit{green} (low). A human operator uses this data to assess safety, while autonomous patrolling reduces human exposure to harmful radiation.

If the rover detects high or medium radiation, it returns to the entry point (waypoint 0) for decontamination, preventing damage and avoiding human retrieval in hazardous conditions.

The rover’s high-level requirements are:
\begin{description}
\item[REQ1:] Inspect each waypoint unless radiation is high.
\item[REQ2:] Inspect each waypoint at least once.
\item[REQ3:] Abort and return to entry on high radiation.
\item[REQ4:] Complete all inspections or abort due to high radiation.
\end{description}

The system comprises four components: \textbf{Localisation}, which estimates the rover’s position; \textbf{Navigation}, which handles path planning and movement tracking; the \textbf{Agent}, which makes decisions based on the rover’s position and radiation level (\textit{radiationStatus}); and the \textbf{Radiation Sensor}, which monitors radiation and confirms inspections. Built using \gls{ros}, the system supports trace collection via a global clock.\footnote{\url{https://wiki.ros.org/Clock}} Future work will explore decentralised \gls{rv}.

Our \gls{csp} model captures the minimal behaviour satisfying REQ1--REQ4. We verify it with \gls{fdr} and reuse it directly in \Varanus as the \gls{rv} oracle. Since the \gls{sua} may perform additional, irrelevant events, we use \Varanus in permissive mode, requiring only that events from the model’s alphabet occur in the specified order and with correct parameters.

\begin{sloppypar}
Fig.~\ref{fig:roverCSP} shows the main \gls{csp} processes. If $WaypointSet = \emptyset$, $ROVER(\emptyset, \_)$ ends the mission. Otherwise, $ROVER(WaypointSet, Rad)$ forms the main loop: the rover issues an $inspect$ command for a waypoint $wp \in WaypointSet$, then transitions to $ROVER\_INSPECTING(WaypointSet, Rad, wp)$ to perform $move.wp$. After moving, it returns to $ROVER$ with $wp$ removed ($diff(WaypointSet, \{wp\})$).
\end{sloppypar}

A guarded $move.0$ event represents entry into the store and is only enabled if 0 is in $WaypointSet$. The $radiation\_level$ event is always available. If the level is $Red$ or $Orange$, the process transitions to $ROVER\_ABORT$, which triggers $move.0$ and aborts the mission. If the level is $Green$, then the rover continues with the $Rad$ parameter updated to match.

\begin{figure}[t]
    \centering
 
\begin{syntax}
ROVER(\emptyset, \_) = \\
\quad mission\_complete \then \Skip\\
\quad \extchoice \\
\quad 	radiation\_level?Green \then ROVER(\emptyset, Green) \\
\quad \extchoice \\
\quad	radiation\_level?r:({Red, Orange}) \then ROVER\_ABORT \\
~\\
ROVER(WaypointSet, Rad) =\\ 	
\quad	inspect?wp:(WaypointSet) \then ROVER\_INSPECTING(WaypointSet, Rad, wp)\\
\quad 	\extchoice \\
\quad	radiation\_level?Green \then  ROVER(WaypointSet, Green)\\
\quad	\extchoice \\
\quad	radiation\_level?r:({Red, Orange}) \then ROVER\_ABORT \\
\quad	\extchoice \\
\quad 	(member(0,WaypointSet)) \& move.0 \then ROVER(diff(WaypointSet, {0}), Rad)\\
~\\
ROVER\_INSPECTING(WaypointSet, Rad, wp) = 	\\
\quad	move.wp \then ROVER(diff(WaypointSet, {wp}), Rad)	\\
\quad \extchoice \\
\quad	radiation\_level?Green \then ROVER\_INSPECTING(WaypointSet, Green, wp)\\
\quad	\extchoice \\
\quad	radiation\_level?r:({Red, Orange}) \then ROVER\_ABORT\\
~\\
ROVER\_ABORT = move.0 \then mission\_abort \then \Skip
\end{syntax}

\caption{Specification of the rover's abstract behaviour. Using pattern-matching $ROVER(\emptyset, \_)$ defines the processes behaviour when the $WaypointSet$ is empty (in CSP $\_$ is a wildcard) otherwise the process behaves as $ROVER(WaypointSet, Rad)$.} 
\label{fig:roverCSP}
\end{figure}

The specification in Fig.~\ref{fig:roverCSP} satisfies REQ1--REQ4. It checks radiation before inspection (REQ1), removes inspected waypoints from the set (REQ2), returns to waypoint 0 on $Red$ or $Orange$ readings (REQ3), and terminates only when all waypoints are inspected or high radiation is detected (REQ4).

\begin{sloppypar}
The model begins with $mission\_start \then ROVER(waypointID, Green)$, where $waypointID$ is the full set of waypoints and $Green$ is assumed at the entry point. Upon termination via $mission\_complete$ or $mission\_abort$, a new mission can begin. The events $mission\_start$, $mission\_complete$, and $mission\_abort$ help identify mission boundaries in traces.
\end{sloppypar}

\begin{sloppypar}
\Varanus synthesises an \gls{lts} from this model and monitors traces as described in \S\ref{sec:varanusOverview}. For example, the trace $\langle mission\_start, inspect.2, radiation\_level\allowbreak.Green, move.2,~\ldots \rangle$ passes, as $inspect$ and $move$ target the same waypoint and radiation is safe. In contrast, the trace $\langle mission\_start, inspect.2, radiation\_level\allowbreak.Red, move.2,~\ldots \rangle$ fails, since the rover should return to waypoint 0 after detecting high radiation. In the next section, we evaluate \Varanus’s performance and its ability to catch such violations.
\end{sloppypar}

\section{Evaluation}
\label{sec:evaluation}

We evaluate \Varanus via two experiments. First, we stress-test it using generated models and traces of increasing size. Then, we demonstrate its effectiveness by monitoring event traces from the simulated rover case study -- first using a passing trace, then injecting faults. While \Varanus supports online \gls{rv}, instrumenting this particular case study proved challenging, so making it easier to map \gls{ros} messages into events in the model is left for future work.

\subsection{Stress Testing}
\label{sec:offlineEval}

We stress-test \Varanus’s performance during \gls{lts} synthesis and verification by using increasingly large \gls{csp} models and traces. 
Our experiments were run on a machine with an Intel Core i7 2.80 GHz 4-core CPU and 16~GiB RAM, and record the timings using the \texttt{time} module in Python's standard library.

\begin{figure*}
    \centering
    \begin{minipage}[b]{0.5\textwidth}
        \centering
        \includegraphics[width=\textwidth]{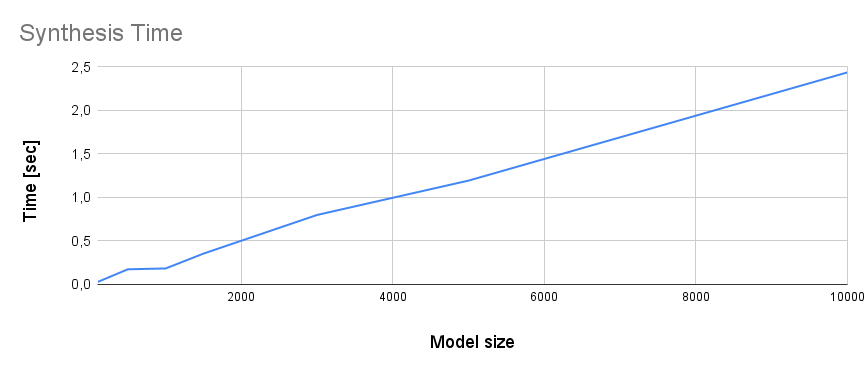}
        \caption{Monitor synthesis time.}
        \label{fig:synthesis}
    \end{minipage}%
    \hfill
    \begin{minipage}[b]{0.5\textwidth}
        \centering
        \includegraphics[width=\textwidth]{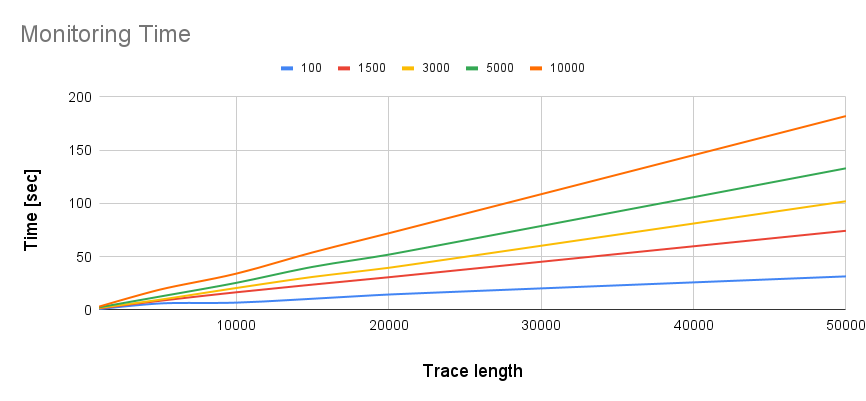}
        \caption{Verification time (whole trace).}
        \label{fig:verification-whole-trace}
    \end{minipage}%
    \hfill
    \begin{minipage}[b]{0.5\textwidth}
        \centering
        \includegraphics[width=\textwidth]{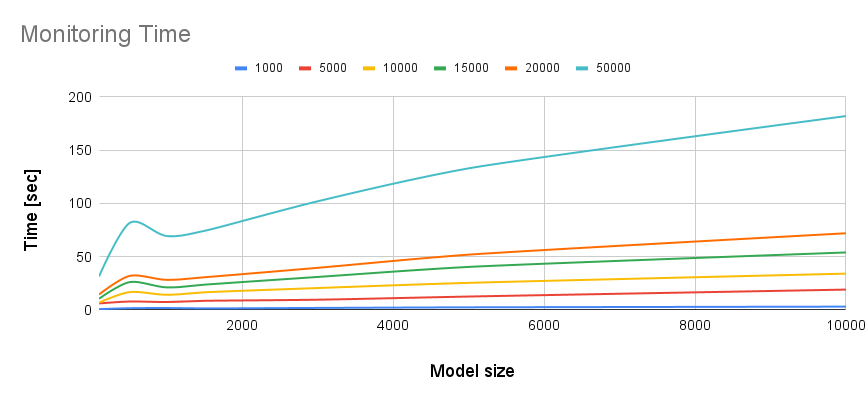}
        \caption{Verification time (whole trace) but varying the model size.}
        \label{fig:verification-whole-trace-model}
    \end{minipage}%
    \hfill
    \begin{minipage}[b]{0.5\textwidth}
        \centering
        \includegraphics[width=\textwidth]{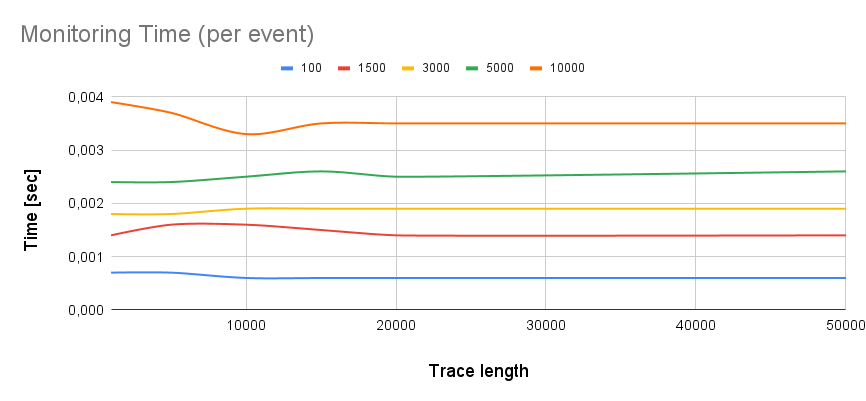}
        \caption{Verification time (single event).}
        \label{fig:verification-single-event}
    \end{minipage}%
    \hfill

   \begin{minipage}[b]{0.5\textwidth}
        \centering
        \includegraphics[width=\textwidth]{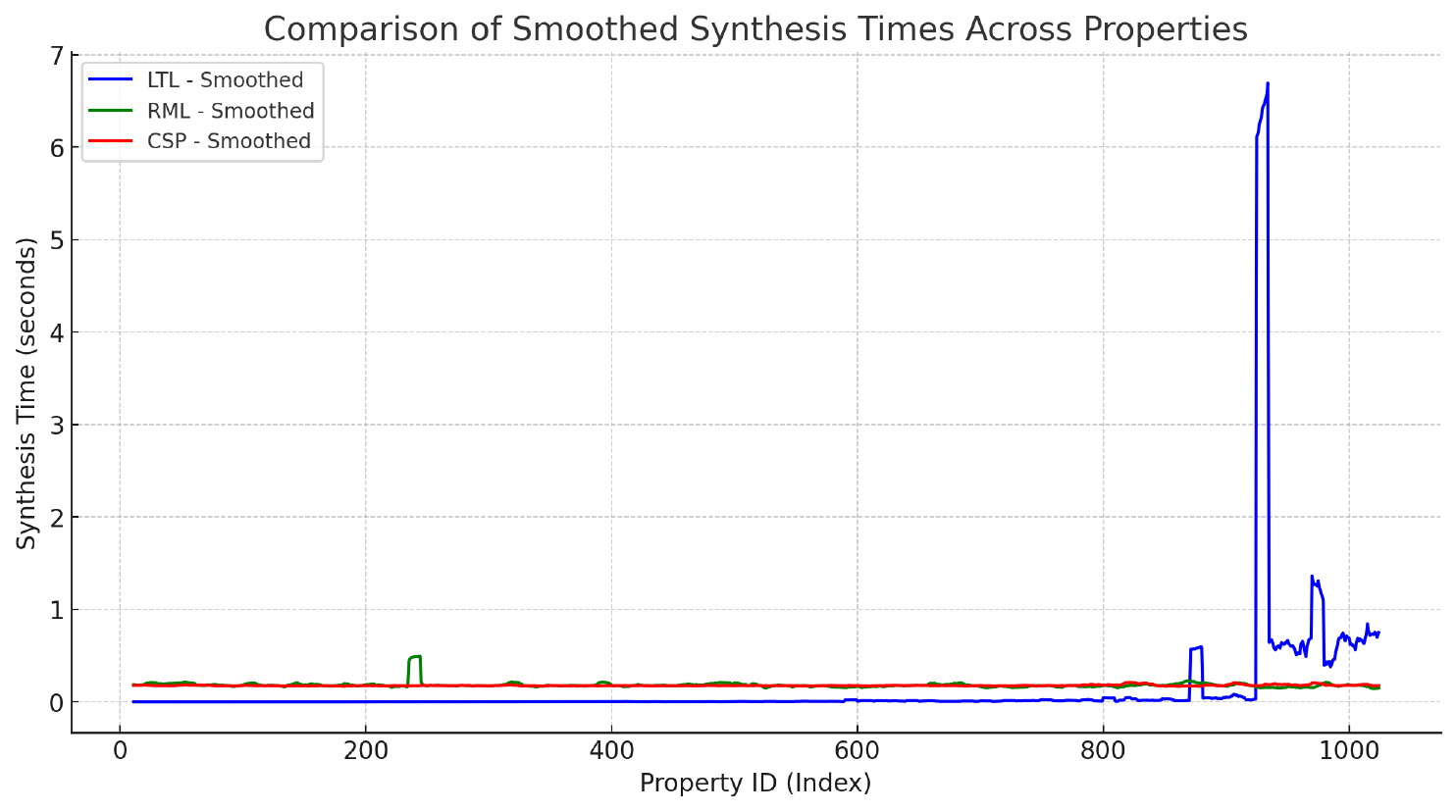}
        \caption{Comparison of Synthesis Time for LTL, RML, and (our synthesis of) CSP. The Property ID is the identifier of the LTL properties analyses, in which higher is the ID, bigger is the size of the LTL property.}
        \label{fig:comparison}
   \end{minipage}

\end{figure*}

To create a `worst-case process', we generate \gls{csp} models of size $N$ such that the corresponding \gls{lts} has $N$ states, each with transitions to all others -- e.g., a model of size 10 yields 10 states and 100 transitions. These maximal-transition models subsume any other finite \gls{lts} of the same size and alphabet.
Fig.~\ref{fig:synthesis} shows that synthesis time grows roughly linearly with model size. Each data point represents the average of 10 runs.

We generate traces of varying lengths, using the alphabet of the \gls{csp} models from the previous test, to evaluate how \Varanus's performance varies by trace length. Any generated trace is valid, since the \gls{csp} models always accept any event in the alphabet. 
Fig.~\ref{fig:verification-whole-trace} shows the time to verify complete traces for models of different sizes. The x-axis represents trace length; the y-axis, verification time. Each line corresponds to a model size (e.g., 100 to 10,000 states). As expected, verification time increases linearly with trace length, indicating \Varanus’s suitability for online \gls{rv}.
Fig.~\ref{fig:verification-whole-trace-model} reorganises the same data, here the lines corresponds to a different trace length. The results show linear scaling with both model size and trace length.

Fig.~\ref{fig:verification-single-event} presents the average time to verify individual events. Unlike previous plots, it highlights that \Varanus can process each event in constant time -- an essential property for online \gls{rv}, where events arrive incrementally.

We also compare \Varanus’s synthesis time with two existing approaches: \gls{ltl} via the Spot library~\cite{DBLP:conf/cav/Duret-LutzRCRAS22} and \gls{rml} via SWI-Prolog~\cite{DBLP:journals/tplp/WielemakerSTL12}. Fig.~\ref{fig:comparison} shows synthesis times for over 1,000 randomly generated \gls{ltl} formulae, with equivalent \gls{csp} and \gls{rml} specifications derived from each\footnote{Restricted to the safety fragment of \gls{ltl} to enable translation, following~\cite{DBLP:conf/fm/LeuschelMC01}.}. \gls{rml}~\cite{DBLP:journals/scp/AnconaFFM21} is included due to its process algebra roots and conceptual similarities to \Varanus.
As Fig.~\ref{fig:comparison} shows, \gls{ltl} synthesis time degrades with formula size, but both \Varanus and \gls{rml} maintain stable, scalable performance.

\subsection{Monitoring the Case Study}
\label{sec:onlineEval}

To evaluate \Varanus on the simulated rover from \S\ref{sec:caseStudy}, we collected a 243-event trace from a run where the rover patrolled five waypoints with all radiation levels reported as $Green$. We included $mission\_start$ and $mission\_complete$ events to mark the trace boundaries. The \textbf{Radiation Sensor} in the simulation continuously emits $radiation\_level$ events, which the model accommodates.

These experiments assess whether \Varanus produces correct verdicts, rather than focusing on synthesis or verification time. Table~\ref{tab:caseStudyTimes} summarises performance metrics. We ran our experiments on a machine with an Intel Xeon 3.50GHz 8-core CPU and 32~GiB RAM on Ubuntu~22.04.4.

\begin{table*}[t]
\caption{Summary of approximate total time, synthesis time, checking time, and mean checking time per event, for the four experiments in \S\ref{sec:onlineEval}. The final column shows the number of events that were checked and the total number of events in the trace, for each experiment, and if the trace passed or failed.}

\centering
{

\begin{tabular}{l|l|l|l|l}
 Total Time& Synthesis Time  & Checking Time  & Mean Time/Event & Events/Total Events   \\ \hline \hline
0.39s           & 0.22s              & 0.14s              & 0.0004s                      & 243/243 (Pass)                \\ \hline
 0.27s          & 0.19s              & 0.05s              & 0.0004s                      & 51/243 (Fail)                 \\ \hline
 0.30s           & 0.22s               & 0.05s              & 0.0004s                      & 49/243 (Fail)                 \\ \hline
0.31s           & 0.20s               & 0.08s              & 0.0004s                      & 121/243 (Fail)                \\ \hline
\end{tabular}%

}

\label{tab:caseStudyTimes}
\end{table*}

Running \Varanus on the full 243-event trace yields no violations, completing in $\sim$0.39s. This is expected, as all events are correctly ordered and radiation levels remain $Green$. The \gls{lts} synthesis took $\sim$0.22s, and trace checking $\sim$0.14s, with an average of 0.0004s per event.
To test failure detection, we inject a fault where the rover encounters $radiation\_level.Red$ but continues the mission, violating REQ3. \Varanus aborts after 51 events, correctly expecting $move.0$, and completes in $\sim$0.27s (synthesis: $\sim$0.19s; checking: $\sim$0.05s).

We also evaluate \Varanus’s ability to detect incorrect event ordering and parameter mismatches. Reversing $inspect.1$ and $move.1$ causes a failure after 49 events (total time: $\sim$0.30s; synthesis: $\sim$0.22s; checking: $\sim$0.05s). Injecting a mismatch (e.g., $inspect.3$ followed by $move.5$) causes \Varanus to abort after 121 events, completing in $\sim$0.31s (synthesis: $\sim$0.20s; checking: $\sim$0.08s). In all cases, the average checking time remains $\sim$0.0004s per event.

As shown in Table~\ref{tab:caseStudyTimes}, the mean checking time per event is $\sim$0.0004s, even for the full passing trace. Fig.~\ref{fig:verification-single-event} shows that even under pessimistic conditions, this time remains below 0.004s for models with up to 10,000 states and 50,000 events.
These results suggest that \Varanus is suitable for online \gls{rv}, with minimal runtime overhead. Although \gls{lts} synthesis is relatively slow, it occurs only once before monitoring begins and does not affect runtime performance. Future work aims to improve synthesis speed (\S\ref{sec:conclusion}). \Varanus effectively detects deviations from the \gls{csp} model, including violations of system requirements and incorrect event ordering or parameter mismatches.

\section{Conclusion and Future Work}
\label{sec:conclusion}

This paper presents \Varanus, a novel \gls{rv} tool that verifies a system’s behaviour against a \gls{csp} specification. \Varanus builds an \gls{lts} from a \gls{csp} process and uses it as an oracle to validate the \gls{sua}'s actions. While the specification is parsed using \gls{fdr}, event checking is performed independently on the synthesised \gls{lts}.

We demonstrated \Varanus using a simulated autonomous rover and stress-tested it with worst-case models and traces of increasing size. \Varanus synthesises the \gls{lts} it uses as its oracle in roughly linear time and checks each event in constant time. In the rover case study, the mean event checking time was $\sim$0.0004s, indicating that \Varanus introduces minimal overhead and is suitable for online \gls{rv}. Reusing unmodified \gls{csp} specifications simplifies the modelling process, which is often the primary bottleneck in applying formal methods~\cite{rozier2016}.

\gls{csp} enables the specification of both high-level requirements and detailed designs, linked by refinement, and is sensitive to event order -- essential for message-based systems.
This makes it particularly useful for \gls{rv} of RAS, because it bridges the gap between design-time models and dynamic real-world behaviour.  \Varanus also enhances the \gls{csp} toolset by supporting validation of existing models against real implementations.

In future work, we aim to remove the dependency on \gls{fdr} by developing a stand-alone \gls{csp} parser. This would improve synthesis time and enable support for platforms where \gls{fdr} is unavailable, such as ARM-based systems. We also plan to enable runtime \textit{enforcement}, where incorrect behaviours are corrected by selecting valid alternatives from the model. This requires system architectures that consult \Varanus before executing behaviours~\cite{luckcuck_using_2021}, as well as support for replanning or automated selection of valid next steps from counterexamples. To facilitate this, we will link \Varanus to the ROSMonitoring framework~\cite{FerrandoROSMon2020}, which will also support future demonstrations of online \gls{rv}.

We further intend to demonstrate \Varanus on more complex examples, including larger models and scenarios with multiple monitors (e.g., one per component). This will allow us to explore the impact on synthesis and checking time at scale. Additionally, we are interested in using \Varanus for predictive \gls{rv}, where the \gls{csp} model is used to detect behaviours that inevitably lead to failure. Combined with multiple monitors, this opens the door to multi-monitor predictive \gls{rv}~\cite{FerrandoBriding2021}. Finally, we plan to use \Varanus to monitor \gls{csp} mission specifications, ensuring robotic systems adhere to their intended missions~\cite{MacConvilleCSP2Turle2023}.

\section*{Acknowledgements}
Some of Luckcuck's initial work was done while working for the Universities of Liverpool, Manchester, and Maynooth. We are grateful to Faruq for her work on this paper \textit{in her spare time}, and to Pedro Ribeiro for his help fixing FDR and his general interest in the work.

\bibliographystyle{plain}
\bibliography{CSP-Monitoring.bib} 

\end{document}